\documentclass[sigconf]{acmart}

\setcopyright{acmlicensed}
\copyrightyear{2026}
\acmYear{2026}
\acmDOI{XXXXXXX.XXXXXXX}

 \acmConference[ASE '26]{41st IEEE/ACM International Conference on Automated Software Engineering}{October 12--16, 2026}{Munich, Germany}

\usepackage{subcaption}
\usepackage{todonotes}
\usepackage{multirow}
\usepackage[most]{tcolorbox}
\usepackage{amsmath}
\usepackage{tcolorbox}
\usepackage{tabularx}
\usepackage{booktabs}
\usepackage{multirow}

\begin{document}

\title{Bridging Stakeholder and Product Requirements: An Empirical Study of Requirement Engineering in the Automotive Industry}

\author{Zixu Wang}
\email{zixu.wang@tum.de}
\orcid{0009-0002-3951-0131}
\affiliation{
  \institution{Technical University of Munich, 
  Infineon Technologies AG}
  \city{Munich}
  \country{Germany}
}

\author{Shengcheng Yu}
\email{shengcheng.yu@tum.de}
\orcid{0000-0003-4640-8637}
\authornote{Shengcheng Yu is the corresponding author.}
\affiliation{
  \institution{Technical University of Munich}
  \city{Heilbronn}
  \country{Germany}
}

\author{Zhenchang Xing}
\email{zhenchang.xing@data61.csiro.au}
\orcid{0000-0001-7663-1421}
\affiliation{
  \institution{CSIRO’s Data61}
  \country{Australia}
}

\author{Tobias Wenzel}
\email{tobias.wenzel@infineon.com}
\orcid{0009-0004-6660-4816}
\affiliation{
  \institution{Infineon Technologies AG}
  \city{Munich}
  \country{Germany}
}

\author{Chunyang Chen}
\email{chun-yang.chen@tum.de}
\orcid{0000-0003-2011-9618}
\affiliation{
  \institution{Technical University of Munich}
  \city{Heilbronn}
  \country{Germany}
}

\begin{abstract}

The automotive industry's shift toward software-driven systems has increased complexity while raising the stakes for requirement intake and refinement—critical not only for correctness and compliance, but also for development speed and systematic reuse. While prior research has proposed techniques for improving requirement quality, there is limited empirical understanding of how stakeholder-level requirements are evaluated, refined, and transformed into product-level requirements in industrial automotive practice. This paper presents a large-scale empirical study of requirements engineering based on an industrial dataset from Infineon comprising 8,082 stakeholder requirements and 5,870 product requirements, enriched with traceability links, decision outcomes, deviation rationales, and domain references. Using a mixed-methods approach that combines quantitative analyses of requirement structures, decision distributions, and mapping patterns with qualitative analysis of rationales and referenced specifications, and software- and hardware-related artifacts, we investigate structural and contextual differences between stakeholder and product requirements, factors influencing acceptance, rejection, and approval with deviation, and the nature of stakeholder-to-product requirement refinement. The results reveal systematic differences across abstraction levels and show that refinement complexity is driven primarily by architectural scope and missing contextual information rather than linguistic verbosity. We further derive a taxonomy of stakeholder–product requirement mapping patterns and relate them to differing refinement effort. These findings provide concrete insights into industrial requirements intake and refinement practices and highlight actionable opportunities for improving intake validation, deviation management, and tool-supported contextual enrichment to support faster and more reusable automotive product development.

\end{abstract}



\maketitle

\section{Introduction}
\label{chapter/1_introduction}

The automotive industry is undergoing a profound transformation from mechanically centered products toward software-driven and cyber-physical systems~\cite{broy2006challenges}. Modern vehicles increasingly rely on highly integrated software and hardware platforms that coordinate functionality across hundreds of electronic control units interconnected through complex in-vehicle networks~\cite{ng2021robotic, greengard2015automotive, pancik2018auto}. Software now governs not only infotainment and comfort features but also powertrain control, advanced driver assistance systems, and safety-critical functions~\cite{prasad2006case, bergmiller2013design}. As a result, automotive systems exhibit unprecedented scale, heterogeneity, and interdependence across software, hardware, and communication layers~\cite{broy2006challenges, dajsuren2019automotive}. This shift has led to a substantial escalation in system complexity and integration effort~\cite{pretschner2007software, haghighatkhah2017automotive}. The growing volume of software and its tight coupling with hardware have increased the likelihood of integration defects, late discovery of inconsistencies, and unexpected emergent behavior~\cite{mayur2025advancing}. In recent years, a significant proportion of vehicle recalls and field failures have been attributed to software-related issues, including safety-critical defects~\cite{pancik2018auto}. These developments have led to growing concerns in industry regarding the need for rigorous, lifecycle-wide quality assurance.

Quality assurance in the automotive industry is strongly shaped by a constellation of standards and frameworks. Functional safety in road vehicles is governed by ISO 26262~\cite{ISO26262-2018}, which mandates safety-driven refinement and traceable justification of safety requirements. AUTOSAR~\cite{AUTOSAR_Standard} standardizes software architecture, component interfaces, and integration mechanisms, thereby influencing how requirements are decomposed and allocated. Automotive SPICE~\cite{VDAQMC2017_ASPICE_PAM} defines process requirements and assessment models for system and software development, including explicit processes for requirements engineering alongside many other process areas. Together, these standards and specifications demand not only technically correct implementations but also auditable, well-structured, and standards-compliant requirements throughout refinement and realization.

Within this context, requirements engineering (RE) plays a fundamental role in quality assurance~\cite{ISO26262-2018, pohl2016advanced, VDAQMC2017_ASPICE_PAM}. Requirements define intended system behavior, constraints, and operating conditions prior to implementation. Decisions made during early requirements elicitation and refinement have a lasting impact on defect rates, downstream rework, compliance with standards, and ultimately certification success~\cite{boehm1988understanding, nuseibeh2000requirements,arthur2022foundations, damian2005requirements}. Deficiencies at the requirements level often propagate across development phases, becoming increasingly costly and difficult to correct once embedded in architectures and implementations~\cite{damian2005requirements}. RE is not only a means of assuring quality and compliance, but also a prerequisite for efficient product-line engineering, faster requirement intake, and systematic reuse across product variants~\cite{braun2014guiding, weber2002requirements}. Despite its importance, requirements engineering in the automotive industry faces persistent challenges~\cite{dajsuren2019automotive, haghighatkhah2017improving}. Requirements are predominantly specified in natural language, which makes them susceptible to ambiguity, underspecification, and implicit assumptions~\cite{handbook2003contract, nuseibeh2000requirements}. Furthermore, requirements can be managed across multiple abstraction levels, ranging from high-level stakeholder requirements to detailed product and component specifications~\cite{pohl2024fundamentals}. This process spans organizational boundaries among original equipment manufacturers, tier-1 suppliers, and semiconductor vendors. Coordinating these actors while preserving intent, context, and traceability is inherently complex and error-prone~\cite{winkler2010survey}. 

Prior research on general RE has proposed a variety of approaches to mitigate these challenges. Existing work includes quality models~\cite{femmer2017requirements, femmer2017rapid} and defect taxonomies for natural-language requirements~\cite{ott2012defects}, knowledge extraction and representation~\cite{schlutter2020knowledge, schlutter2018knowledge}, test specification techniques~\cite{milchevski2025multi}, automated linguistic checks~\cite{bertram2023leveraging, farfeleder2011dodt}, and traceability recovery methods~\cite{niu2025tvr, fuchss2025lissa}. Tool support has been developed to establish and analyze links among requirements and related artifacts~\cite{schlutter2018knowledge, niu2025tvr, bonner2023automated}. However, most of these approaches focus on assessing the quality of isolated requirement sets or on recovering missing links after the fact. They rarely address the end-to-end refinement process from stakeholder-level requirements to product-level specifications as it unfolds in industrial practice.

Consequently, several critical gaps remain. There is limited industrial\allowbreak-scale evidence on how and why stakeholder requirements are actually evaluated and transformed. The structural forms of stakeholder-to-product requirement mappings, the emergence of complex refinement structures, and the contextual information engineers must reconstruct during refinement are not well understood. In particular, there is a lack of empirical insight into how stakeholder-level and product-level requirements differ structurally and how they are related in real automotive projects.

This paper addresses these gaps through an empirical study based on an industrial dataset from Infineon. The dataset comprises 8,082 stakeholder requirements and 5,870 product requirements. It includes explicit traceability between product requirements and their source stakeholder requirements, documented approval/rejection decisions (with rationales), and references to specifications and auxiliary domain artifacts. To the best of our knowledge, this constitutes one of the most comprehensive datasets available for studying requirements refinement in an industrial automotive setting. With this dataset, the study investigates several interrelated questions. It examines how stakeholder and product requirements differ in structure and contextual richness, which factors influence acceptance, rejection, and approval with deviation decisions, and how stakeholder-to-product requirement mappings are organized. It further analyzes what drives refinement complexity and which types of contextual information engineers reconstruct during the elaboration of product requirements.

This study adopts a mixed-methods approach. Quantitative analyses are used to characterize requirement structures, decision distributions, and mapping patterns through descriptive statistics and structural comparisons. Pattern mining is applied to identify recurring forms of stakeholder-to-product requirement relationships. In parallel, qualitative analyses examine decision rationales, deviation justifications, and references to industry specifications and hardware documentation using thematic coding.

The results reveal systematic structural differences between stakeholder and product requirements and provide empirical evidence on how missing context and specification-driven constraints influence decision outcomes. The study derives a taxonomy of stakeholder-to-product requirement mapping patterns and associates these patterns with differing levels of refinement effort. Importantly, the findings indicate that refinement complexity is driven less by linguistic verbosity and more by architectural scope and implicit domain assumptions. Engineers frequently need to reconstruct context from specifications and hardware documentation to make stakeholder requirements implementable. These insights highlight concrete opportunities to improve requirement intake validation, deviation management, and tool-supported contextual enrichment in industrial automotive requirements engineering.

This paper conducts the first empirical study to jointly analyze stakeholder-level and product-level requirements at an industrial scale. In summary, this paper makes the following noteworthy contributions: 

{\sloppy
\begin{itemize}
\item A first large‑scale empirical characterization of stakeholder‑level and product‑level requirements (SHRQs and PRQs) in an industrial automotive setting, revealing systematic differences in structure, granularity, and contextual completeness across abstraction levels.
\item An analysis of acceptance and deviation decisions, showing how industry specifications, feasibility constraints, and missing context influence requirement outcomes.
\item A taxonomy of SHRQ–PRQ mapping relationships and refinement complexity, demonstrating that complexity is driven primarily by architectural scope and missing contextual information rather than textual verbosity.
\item Implications for practice, including improved intake validation, systematic deviation management, and tool support for contextual enrichment, and opportunities 
for future automation in requirement assessment and refinement.
\end{itemize}}


\section{Background and Motivation} 
\label{chapter/2_background}

\subsection{Requirements in the Automotive Domain}
Requirements Engineering (RE) structures how system behavior is specified, refined, and verified throughout the automotive development process~\cite{pohl1996requirements, zave1997four, nuseibeh2000requirements, weber2002requirements, ISO26262-2018, ISO26262-6-2018}. Within this process, requirements are organized across multiple abstraction levels to accommodate the separation of responsibilities among OEMs, tier-1 suppliers, and semiconductor providers, as well as the technical constraints imposed by safety standards and architectural specifications. Understanding these abstraction levels is essential for analyzing how high-level intents are ultimately realized in component-level software behavior~\cite{weber2002requirements, pohl2016requirements}.


\textbf{Stakeholder requirements (SHRQs)} express vehicle-level goals, functional expectations, safety intents, and regulatory constraints. They are typically written in natural language and vary widely in granularity, scope, and completeness. SHRQs capture what the system should achieve from an end-user or safety perspective but often omit timing assumptions, operational modes, and hardware dependencies.

\textbf{Product requirements (PRQs)} refine high-level intents into implementable specifications for specific hardware or software components. They must comply with architectural constraints such as component interfaces, signal models, and timing behavior, as well as platform capabilities and safety requirements derived from ISO 26262. Accordingly, PRQs use controlled syntax and semantics, follow naming conventions, and are organized into functional, interface, performance, and design-constraint categories. To support compliance, PRQs must be atomic, unambiguous, and traceable to both upstream SHRQs and downstream validation artifacts.

\begin{figure}[!t]
\centering
\includegraphics[width=\linewidth]{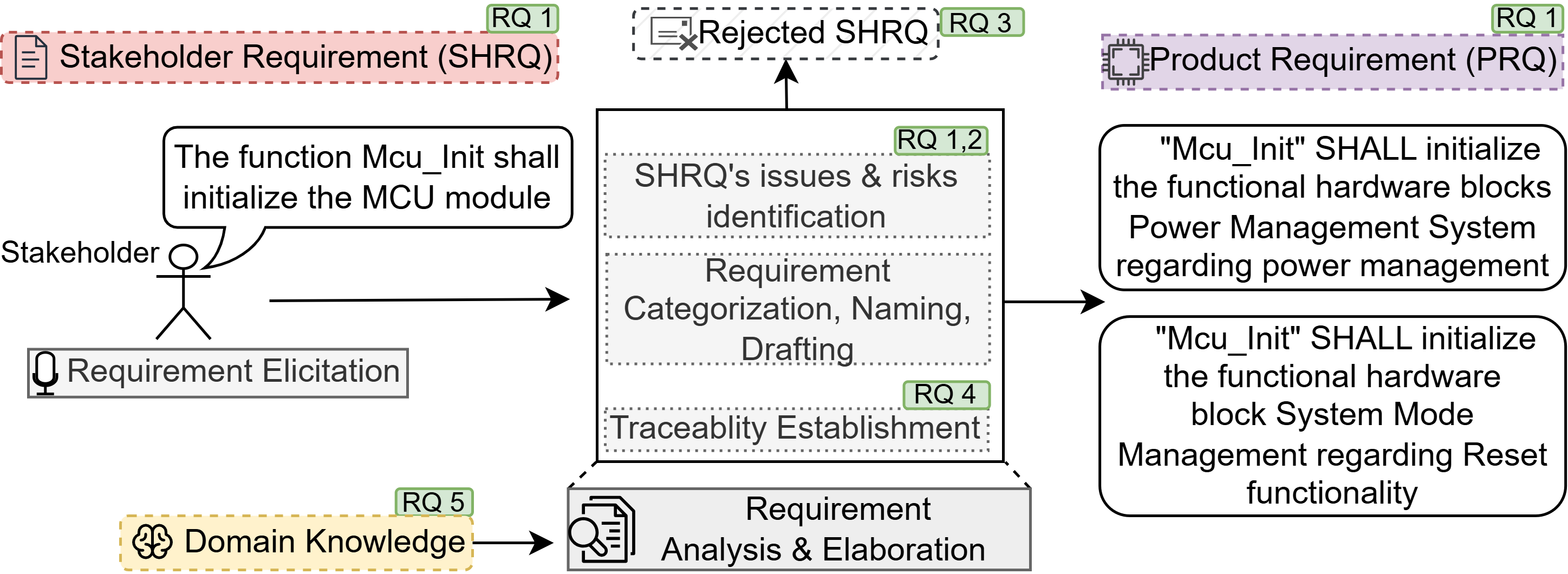}
\vspace{-0.6cm}
\caption{Transformation Process from Stakeholder to Product Requirements in Automotive RE}
\label{fig:Requirement Engineering Process}
\vspace{-0.5cm}
\end{figure}

The refinement from SHRQs to PRQs is a core part of the automotive requirements lifecycle. SHRQs are analyzed, decomposed, and operationalized into increasingly detailed specifications (Fig. \ref{fig:Requirement Engineering Process}) that reflect architectural and hardware constraints. SHRQ–PRQ relationships are often heterogeneous: a single SHRQ may decompose into multiple PRQs covering different components or operational contexts, while multiple SHRQs may converge into one PRQ representing shared functionality. These patterns show that automotive requirements differ not only in abstraction level but also in structure, semantics, contextual completeness, and architectural grounding. This layered organization motivates our subsequent analysis of the challenges and decision mechanisms involved in SHRQ-to-PRQ refinement.



\subsection{Challenges and Motivation} \label{sec: Challenges and Motivation}

Although automotive requirements are organized across abstraction levels, refining SHRQs into PRQs remains difficult to perform consistently in practice. Several factors hinder engineers' ability to operationalize high-level intents while maintaining alignment with safety, architectural, and implementation constraints.

First, SHRQs often lack critical contextual information needed for precise refinement. They frequently omit timing assumptions, operational modes, hardware dependencies, and environmental conditions. For example, a requirement such as ``The system shall support dynamic power-mode transitions'' does not specify transition timing, ECU involvement, or hardware limits. In practice, this led to more than a dozen PRQs across multiple components, including conflicting timing constraints. Engineers must reconstruct missing context from standards, hardware documentation, or internal guidelines, making refinement dependent on tacit knowledge and prone to inconsistency.

Second, SHRQs show substantial heterogeneity in structure, scope, and clarity. Originating from OEMs, regulatory bodies, and diverse internal stakeholders, they vary widely in format and level of detail. This variability complicates the identification of refinement boundaries and increases the risk of ambiguous or conflicting interpretations, directly affecting the derivation of precise PRQs.

Third, refinement relationships are complex. Mappings between SHRQs and PRQs are often many-to-many. A single SHRQ may decompose into multiple PRQs reflecting different architectural roles, timing behaviors, or safety constraints, while several SHRQs may converge into a single PRQ capturing shared functionality. This increases cognitive load and complicates traceability, making it harder to ensure that system-level intent is faithfully realized.

Fourth, acceptance and rejection decisions depend on more than linguistic quality. Engineers frequently approve SHRQs with deviation, reinterpret them to fit architectural constraints, or reject them due to feasibility or safety concerns. However, the rationale behind these decisions is rarely analyzed systematically, limiting insight into current refinement practices.

These challenges call for a deeper empirical understanding of differences between SHRQs and PRQs, refinement decision-making, and the contextual knowledge engineers rely on. This study addresses these gaps through an empirical analysis of industrial automotive projects, characterizing requirement properties, acceptance decisions, refinement structures, and contextual dependencies to ground current practice in evidence and identify opportunities to improve requirement intake, refinement workflows, and future automation.

\section{Research Questions} 
\label{chapter/2.5_RQ}

These challenges reveal gaps in our empirical understanding of how SHRQs and PRQs differ, how engineers evaluate and refine them, and what contextual knowledge is required during refinement. To address these gaps, we investigate five research questions.

\textit{Stakeholder Heterogeneity and Requirement Characteristics}

\textbf{RQ1}: How does stakeholder heterogeneity in the automotive ecosystem manifest in systematic linguistic and structural differences in requirements across abstraction levels?



Understanding these differences provides essential context for interpreting subsequent acceptance decisions and refinement behavior across abstraction levels.

\textit{Requirement Acceptance and Review Rationales}

\textbf{RQ2}: Which characteristics distinguish SHRQs that are accepted from those that are rejected during industrial review?

\textbf{RQ3}: What review rationales most frequently underlie rejection or approval with deviation in industrial practice?

Together, these questions examine both the outcomes of acceptance decisions and the underlying issues that shape these decisions.


\textit{Mapping Complexity}

\textbf{RQ4}: How are SHRQ–PRQ mappings structured in industrial practice, and what factors drive refinement complexity reflected in these mapping patterns?

These questions explore the cognitive and structural aspects of refinement, including architectural decomposition, many-to-many relationships, and cross-component interactions.

\textit{Contextual Knowledge in Refinement}

\textbf{RQ5}: What contextual information is typically missing in SHRQs, how do these gaps co-occur, and how do they shape refinement into PRQs?


This question focuses on identifying which types of contextual information are systematically absent from SHRQs and must be made explicit during product-level refinement. 


\section{Study Design} \label{chapter/3_study_design}
To address the RQs, we conducted an empirical study of real-world automotive requirements in collaboration with Infineon. This section describes the design of the study, including the approach, data, and analysis procedures.

\subsection{Data Source and Collection} 

The data comes from Infineon's automotive chip development projects, which are safety-critical and involve multiple stakeholder groups, like OEM customers, standardization bodies, and internal roles such as system engineering, architecture, safety, and quality. As a result, the requirements reflect the heterogeneity and domain constraints typical of large-scale automotive chip development.

The repository contains requirements at multiple abstraction levels, along with ISO 26262-mandated ~\cite{ISO26262-2018} review decisions and traceability information. For this study, we extracted SHRQs, PRQs, their metadata, including acceptance status and reviewer rationale, and the SHRQ–PRQ traceability links created during refinement.

All artifacts were exported directly from the tool and reviewed to remove direct product identifiers and proprietary technical details. Domain-relevant terminology from public standards such as AUTOSAR and ISO 26262 was retained in representative examples to preserve ecological validity. Because the data originates from real industrial projects and captures actual review outcomes and traceability practices, it provides a reliable basis for analyzing requirement characteristics, acceptance decisions, and refinement behavior in practice.


\begin{table}[h!]
\centering
\caption{Overview of the requirement datasets}
\label{table:dataset_overview}
\vspace{-0.4cm}
\resizebox{0.85\columnwidth}{!}{
\begin{tabular}{p{0.24\columnwidth}p{0.33\columnwidth}p{0.33\columnwidth}}

\hline
\textbf{Attribute} & \textbf{SHRQs} & \textbf{PRQs} \\ \hline
\textbf{Total count} & 8,082 & 5,870 \\ \hline
\textbf{Source} & Derived from specifications (3,939) and non-spec (4,143) & Derived from SHRQs (5,392) and auxiliary sources (478) \\ \hline
\textbf{Status} & Approved: 3,688; Rejected: 4,394 & All approved \\ \hline
\textbf{Safety relevance} & Not applicable & Annotated with ASIL (A–D, QM, N/A; per ISO 26262) \\ \hline
\textbf{Requirement type} & Not applicable & Functional, Performance, Interface, Design Constraint, Process, Quality, Application, Testing \\ \hline
\textbf{Textual form} & Natural language or tables; often informal/semi-structured & Natural language; formalized and structured; atomic\\ \hline
\end{tabular}}
\vspace{-3mm}
\end{table}

\subsection{Dataset Overview \& Characteristics}
\label{sect. Dataset Overview}
The dataset comprises two requirement abstraction levels used in automotive chip development: stakeholder requirements (SHRQs) and product requirements (PRQs). 
The dataset (see Table~\ref{table:dataset_overview}) includes 8,082 SHRQs from diverse sources, consisting of 3,939 specification-derived inputs such as AUTOSAR and 4,143 non-specification SHRQs. Each SHRQ includes a short name, textual description, acceptance status, and review rationales. Of these, 3,688 SHRQs were approved for refinement into PRQs, while 4,394 were rejected, typically with reviewer comments explaining the decision. The dataset also contains 5,870 PRQs, representing implementable requirements organized into functional blocks aligned with the software architecture. Among them, 5,392 PRQs trace back to approved SHRQs, while the remaining PRQs capture auxiliary or internal requirements. PRQs further include structured attributes such as requirement type and are annotated with an Automotive Safety Integrity Level (ASIL), ranging from QM to A–D, enabling analysis of safety relevance. The dataset records explicit traceability links between SHRQs and PRQs, enabling analysis of refinement relationships across abstraction levels. These mappings form the basis for analyzing refinement structure and complexity in subsequent sections.

\section{Results}
\label{chapter/4_results}

\subsection{Stakeholder Heterogeneity \& Requirement Characteristics}

\subsubsection{RQ1: Stakeholder Heterogeneity \& Requirement Characteristics}

To answer RQ1, we analyze how differences in stakeholder roles and responsibilities are reflected in systematic linguistic and structural characteristics of SHRQs and PRQs, further distinguishing SHRQs from different sources.


\textit{Length, structural complexity, and readability}
SHRQs exhibit substantially greater variation in length than PRQs. On average, SHRQs contain 43 words (std: 70), whereas PRQs are significantly shorter and more uniform (mean: 19 words, std: 15). Within SHRQs, specification-derived requirements show relatively consistent lengths with moderate variance (mean: 41 words, std: 36), while non-spec SHRQs are highly heterogeneous (mean: 44 words, std: 91), ranging from very short placeholders to lengthy, multi-clause descriptions (max: 2,103 words vs. 755 words). This pronounced variability reflects the heterogeneous origins of SHRQs, which stem from industry specifications (spec-SHRQs), internal engineering teams, OEM inputs, and requirements inherited from previous product generations (non-spec SHRQs). In contrast, PRQs exhibit a constrained and standardized structure, consistent with their role in supporting implementation and integration.

Length alone, however, does not fully capture how stakeholder heterogeneity translates into linguistic complexity. Following established approaches for automated linguistic analysis of requirements, and using the tooling provided by prior work~\cite{ferrari2017pure, femmer2025description}, we examine syntactic and readability metrics to reveal deeper structural differences between requirement types. In Table~\ref{table: Linguistic characteristics of SHRQ and PRQ}, SHRQs contain substantially longer sentences and nearly twice as many clauses per sentence as PRQs (50.41 vs.\ 19.39 tokens per sentence; 4.69 vs.\ 2.14 clauses). Non-spec SHRQs are consistently the most syntactically demanding, with longer sentences (57.93 vs.\ 44.90 tokens) and more clauses (6.00 vs.\ 3.70) than specification-derived SHRQs. Readability metrics follow the same pattern: SHRQs reach substantially higher Flesch–Kincaid Grade Levels than PRQs (23.56 vs.\ 13.40), with non-spec SHRQs again exhibiting the highest values (26.08 vs.\ 21.68). Together, these results show that stakeholder heterogeneity manifests not only in the scale and variability of requirements but also in their syntactic structure and cognitive complexity.

\begin{table}[t]
\centering
\caption{Linguistic characteristics of SHRQs and PRQs}
\vspace{-0.3cm}
\label{table: Linguistic characteristics of SHRQ and PRQ}
\resizebox{\columnwidth}{!}{
\begin{tabular}{lcccccc}
\toprule
\multirow{2}{*}{Metrics} & \multicolumn{5}{c}{SHRQs} & \multirow{2}{*}{PRQs} \\
\cmidrule(lr){2-6}
& Spec. & Non-Spec. & Accepted & Rejected & Overall & \\
\midrule
\textbf{General} &&&& \\
\quad Number of Rows & 3,939 & 4,143 & 3,688 & 4,394 & 8,082 & 5,870 \\
\quad Unique Sentences & 3,800 & 2,812 & 3,603 & 3,015 & 6,612 & 5,797 \\
\midrule
\textbf{Lexical Complexity} &&&& \\
\quad Total Tokens & 170,405 & 162,909 & 162,175 & 171,267 & 333,314 & 112,417 \\
\quad Lexical Words & 120,413 & 105,322 & 113,992 & 111,819 & 225,735 & 74,384 \\
\quad Vocab (Tok.) & 6,515 & 6,272 & 6,940 & 6,758 & 10,786 & 5,045 \\
\quad Vocab (Lex.) & 6,255 & 6,039 & 6,658 & 6,531 & 10,457 & 4,836 \\
\quad Vocab (Stem) & 5,252 & 4,604 & 5,561 & 5,052 & 8,612 & 4,008 \\
\quad Lexical Diversity & 0.04 & 0.04 & 0.05 & 0.05 & 0.04 & 0.05 \\
\quad Lexical Density & 0.71 & 0.65 & 0.70 & 0.65 & 0.68 & 0.66 \\
\midrule
\textbf{Syntactic Complexity} &&&& \\
\quad Avg. Sent. Len. (Tok.) & 44.90 & 57.93 & 45.01 & 56.80 & 50.41 & 19.39 \\
\quad Avg. Sent. Len. (Lex.) & 31.69 & 37.45 & 31.64 & 37.09 & 34.14 & 12.83 \\
\quad Avg.  \# Clauses & 3.70 & 6.00 & 3.82 & 5.71 & 4.69 & 2.14 \\
\quad Avg. Parse Tree Depth & 6.30 & 5.64 & 6.22 & 5.78 & 6.02 & 4.74 \\
\midrule
\textbf{Readability} &&&& \\
\quad Flesch–Kincaid (Grade) & 21.68 & 26.08 & 21.55 & 25.92 & 23.56 & 13.40 \\
\bottomrule
\end{tabular}}
\vspace{-0.4cm}
\end{table}

\textit{Semantic contrasts.}
Vocabulary analysis reveals two layers of difference between spec- and non-spec SHRQs. At the foundational level, both groups share 
common software engineering terminology such as \textit{function}, 
\textit{configuration}, and \textit{module}, though specification requirement uses these 
terms more frequently, reflecting its emphasis on modular architecture. 
At the domain-specific level, clear divergence emerges: spec-SHRQs are characterized by configuration management vocabulary 
(\textit{variant}, \textit{build}, \textit{multiplicity}, \textit{scope}) 
and error detection (\textit{error}, 
\textit{raise}), whereas non-spec requirements 
center on runtime safety mechanisms (\textit{safety}, \textit{reset}, 
\textit{mechanism}), hardware-level interactions (\textit{register}, \textit{hardware}), and fault response strategies (\textit{reaction}, 
\textit{alarm}). Semantic clustering confirms this 
pattern: 65\% of requirements share a common vocabulary core, 
while domain-specific concerns form distinct, near-pure clusters. At the product level, our analysis reveals a hierarchical refinement pattern with moderate semantic overlap between SHRQs and PRQs. The SHRQ-PRQ relationship thus reflects a refinement hierarchy where lower-level requirements elaborate on higher-level specifications using overlapping vocabulary, while maintaining distinct concerns for safety assurance (SHRQ) versus implementation details (PRQ).

Taken together, these results show that stakeholder heterogeneity in the automotive ecosystem manifests in clear and consistent linguistic and structural differences across sources and abstraction levels. SHRQs are longer, more variable, and linguistically more complex, with non-spec SHRQs exhibiting the greatest heterogeneity. PRQs are substantially shorter, more uniform, and focused on technical and implementation-oriented content.

\begin{tcolorbox}
\textbf{Finding 1:} 
Stakeholder heterogeneity in the automotive ecosystem manifests in systematic linguistic and structural differences across sources and abstraction levels. SHRQs are longer, more variable, and linguistically more complex—especially those originating from non-spec sources, whereas PRQs are shorter, more uniform, and focused on implementation and integration concerns.
\end{tcolorbox}

\subsection{Requirement Acceptance \& Review Rationales}

\subsubsection{RQ2: Features of Approved vs. Rejected SHRQs}
To address RQ2, we analyze the acceptance outcomes of SHRQs and examine how they relate to requirement sources and linguistic characteristics. Acceptance outcomes differ markedly between specification-derived and non-specification-derived requirements. Among the 3,939 spec-SHRQs, 83\% were approved, and 17\% were rejected, whereas non-spec SHRQs exhibit a substantially lower acceptance rate, with approximately 10\% approved and 90\% rejected. This contrast indicates that requirement source and standardization play a dominant role in industrial acceptance decisions.

In contrast, textual characteristics alone are insufficient to explain acceptance outcomes. 
In Table~\ref{table: Linguistic characteristics of SHRQ and PRQ}, while accepted SHRQs tend to exhibit higher lexical density, shorter sentences, and fewer clauses on average, rejected SHRQs display greater variability across these metrics. However, none of these linguistic properties decisively distinguishes accepted from rejected requirements in isolation. Taken together, these results suggest that acceptance decisions are not primarily driven by surface linguistic characteristics.

Overall, these findings show that requirement sources dominate acceptance outcomes, while linguistic characteristics alone are insufficient to explain review decisions. To understand why requirements are rejected or conditionally approved, we next examine the explicit review rationales documented by engineers.

\begin{tcolorbox}
\textbf{Finding 2:} 
Acceptance outcomes are primarily driven by alignment with industry specifications and requirement origin, whereas linguistic characteristics alone do not reliably distinguish approved from rejected SHRQs in industrial review.

\end{tcolorbox}


\subsubsection{RQ3: Rationales for Rejected or Deviation-Approved SHRQs}
To address RQ3, we analyzed the documented rationales associated with rejected SHRQs and those approved with deviation. 
Due to the industrial context of the dataset, rationale documentation follows project-specific practices. The available annotations provide sufficient coverage to identify recurring decision patterns, as confirmed through expert validation. Since the dataset lacks predefined rationale categories, several domain experts with experience in automotive RE independently reviewed a sample of rationale comments and jointly defined a taxonomy of rationale types, resulting in 12 categories for rejection and 10 categories for approval with deviation. Using this taxonomy, the remaining rationale comments were automatically categorized using a large language model (GPT-4o) guided by the expert-defined taxonomy, as illustrated by the prompts on our website. We conducted human-in-the-loop validation, where domain experts reviewed a stratified random sample across categories to assess labeling accuracy. Disagreements were resolved by consensus. 
This mixed approach of expert-driven taxonomy and automated classification allowed us to capture the distribution of rationale types at scale while grounding the analysis in domain expertise.

\begin{figure}[h!]
\centering

\includegraphics[width=\columnwidth]{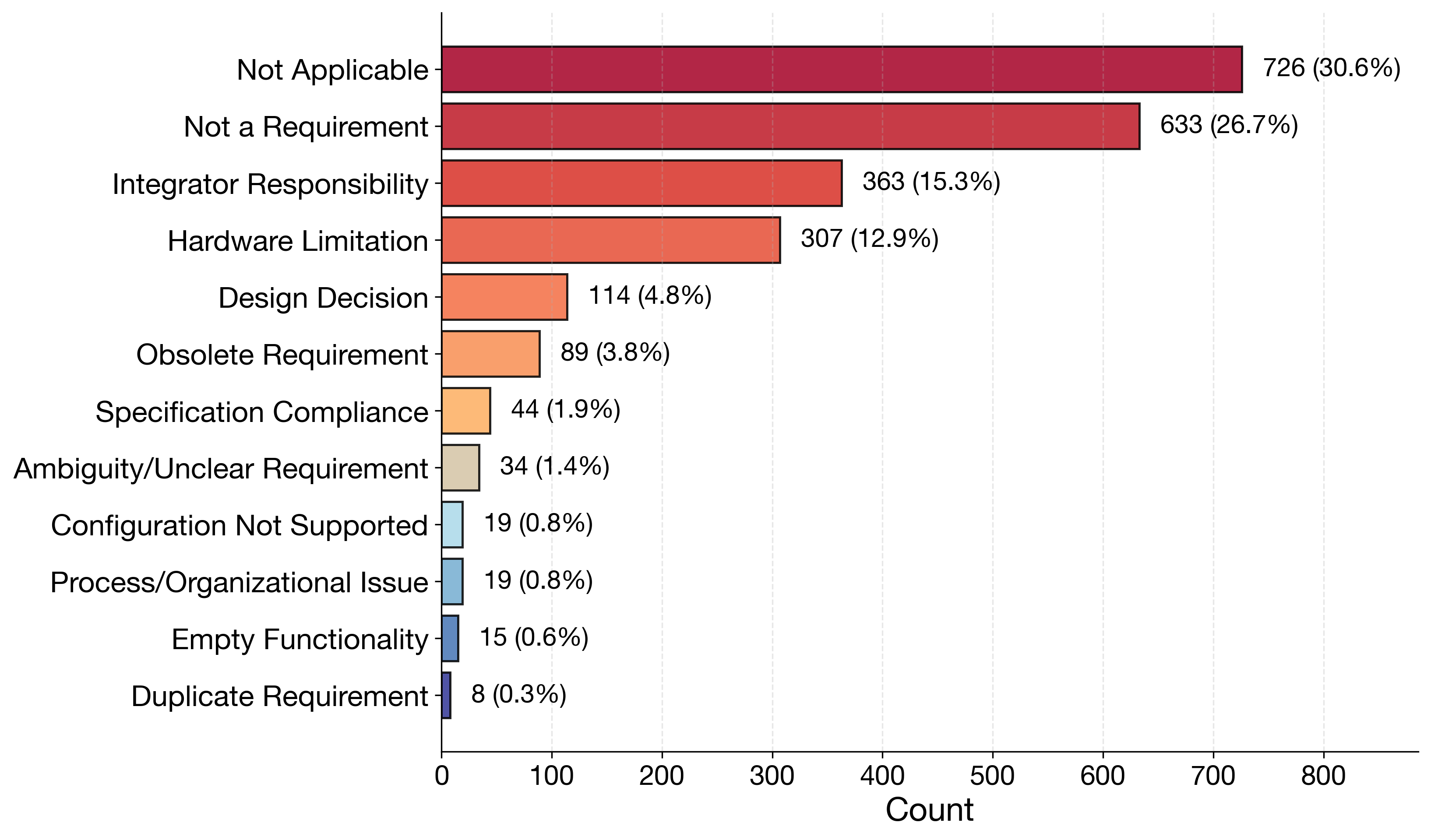}
\vspace{-0.8cm}
\caption{Distribution and Percentage of Rejection Rationale}

\label{fig: Distribution and percentage of rejection rationale}
\end{figure}

\begin{figure}[h!]
\centering
\includegraphics[width=0.9\columnwidth]{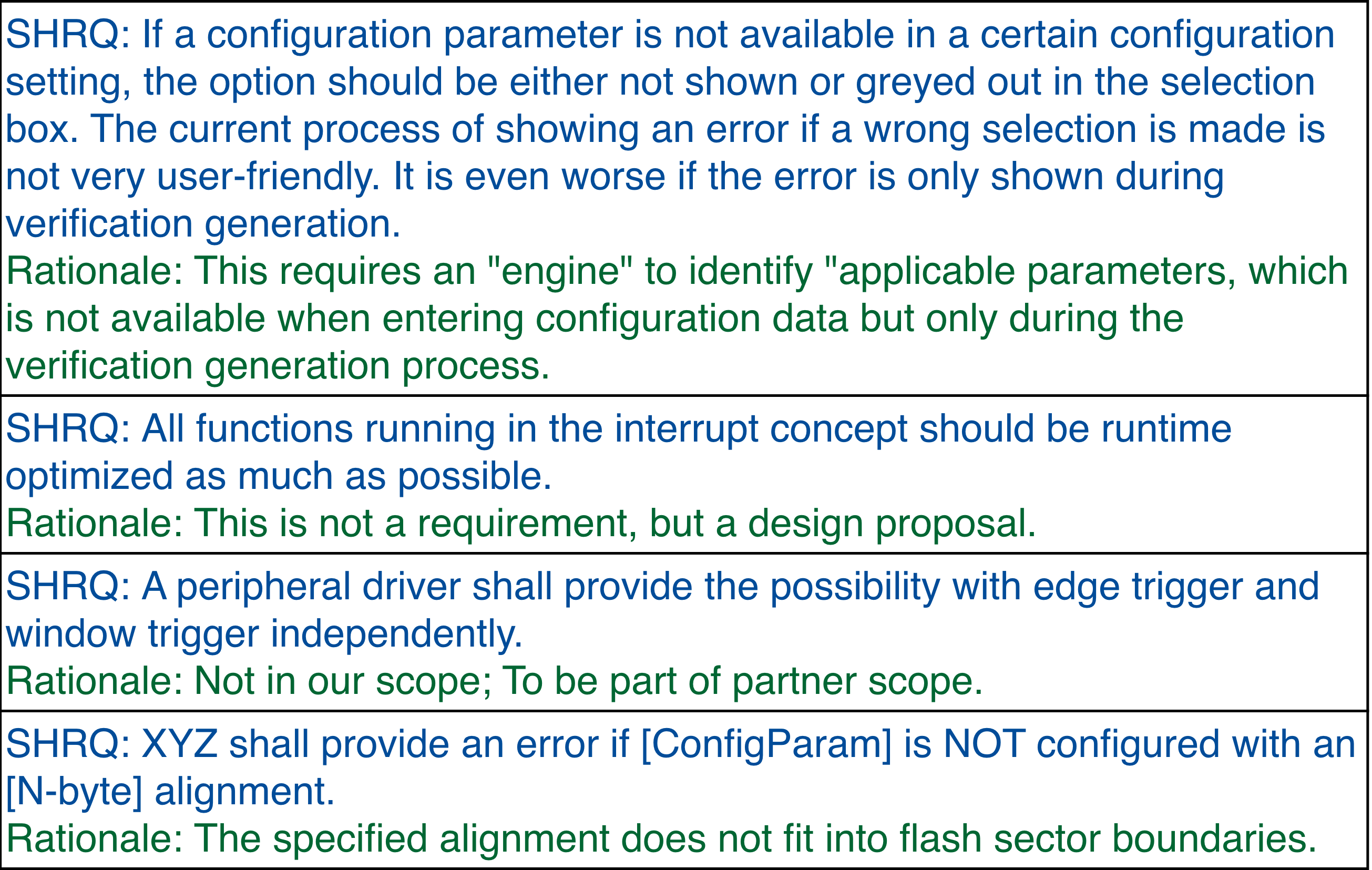}
\vspace{-3mm}
\caption{Examples of Rejection}
\label{fig:Rejection Examples}
\vspace{-3mm}
\end{figure}

\textit{Rejection Rationales.}
The analysis of rejection rationales (Fig.~\ref{fig: Distribution and percentage of rejection rationale}) 
shows that many SHRQs fail because they misalign with the structural and organizational realities of automotive software development. The most frequent causes of rejection are scope- and relevance-related. As shown in Fig.~\ref{fig: Distribution and percentage of rejection rationale}, among the rejected SHRQs, 726 were marked as \textit{Not Applicable}, 633 as \textit{Not a Requirement}, and 363 as the responsibility of higher-level system integrators. Together, these account for over 72\% of rejections with rationale, indicating that a large proportion of stakeholder submissions are not actionable requirements but rather contextual notes, organizational statements, or system-level responsibilities. For example, in Fig.~\ref{fig:Rejection Examples}, one requirement specified that unavailable configuration parameters should be “greyed out” rather than trigger an error, but this was dismissed as \textit{Not Applicable} since it described tooling usability rather than product functionality. Similarly, the second statement was categorized as \textit{Not a Requirement}, because it represented a design suggestion or explanatory note rather than enforceable system requirements. Another common rejection concerned scoping responsibilities: for instance, the third SHRQ in Fig.~\ref{fig:Rejection Examples} was classified as \textit{Integrator Responsibility}, since such functionality belongs to partner systems outside our project’s scope. 
Feasibility constraints represent another recurring rejection factor. \textit{Hardware Limitations} alone account for 12.9\%. The last example in Fig.~\ref{fig:Rejection Examples} shows that one SHRQ demanded an error if a memory configuration parameter was not set to a specific byte alignment, but this was infeasible because the underlying sector boundaries did not support the prescribed alignment constraint. Additional cases were dismissed as \textit{Design Decisions} or \textit{Process-Related} issues, reflecting the frequent mismatch between stakeholder expectations and the realities of embedded hardware and architectures in safety-critical systems. \textit{Specification Compliance} also plays a decisive role. Requirements conflicting with specifications or duplicating existing functionality are typically dismissed, underlining the centrality of alignment with standardized frameworks in automotive software engineering. Finally, intrinsic requirement quality defects, though fewer in number, contribute meaningfully to rejections. \textit{Ambiguity, Unclear} phrasing, \textit{obsolete statements}, or \textit{Empty Functionality} with no actionable semantics undermine engineers’ ability to formalize, implement, or test requirements. While less frequent than scoping issues, these linguistic and semantic defects directly impact the quality of downstream specifications.

\begin{figure}[h!]
\centering

\includegraphics[width=\columnwidth]{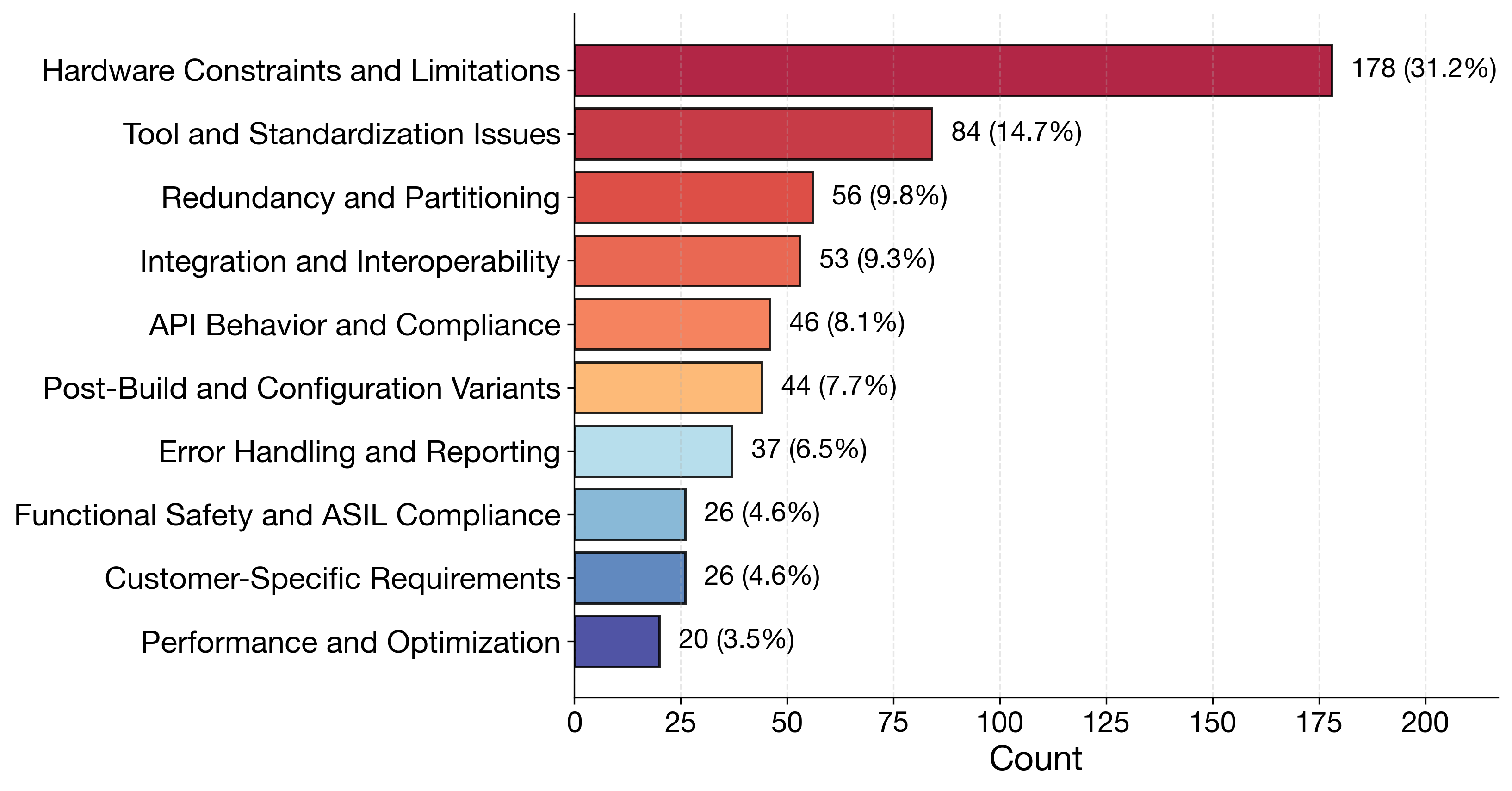}
\vspace{-3mm}
\caption{Distribution and Percentage of Approved with Deviation Rationale}
\label{fig: Distribution and percentage of approved with deviation rationale}
\vspace{-3mm}
\end{figure}

\textit{Approved with Deviation.}
Beyond binary acceptance and rejection, industrial practice also includes a third outcome: \textit{approved with deviation}. Requirements in this category are formally accepted but require modification, clarification, or constraint during refinement. As shown in Fig.~\ref{fig: Distribution and percentage of approved with deviation rationale}, \textit{Hardware Limitations} dominate this category, followed by \textit{Tool or Standardization Issues}, \textit{Redundancy and Partitioning}, and \textit{Integration and Interoperability} challenges. Additional factors include \textit{API Compliance}, \textit{Configuration Variants}, \textit{Error Handling}, and \textit{Functional Safety or ASIL Compliance}. In contrast to rejected SHRQs, which often fail due to scope, deviation cases expose the practical challenges of reconciling even well-specified requirements with hardware, toolchains, and integration realities.

\begin{figure}[h!]
\centering
\includegraphics[width=0.9\columnwidth]{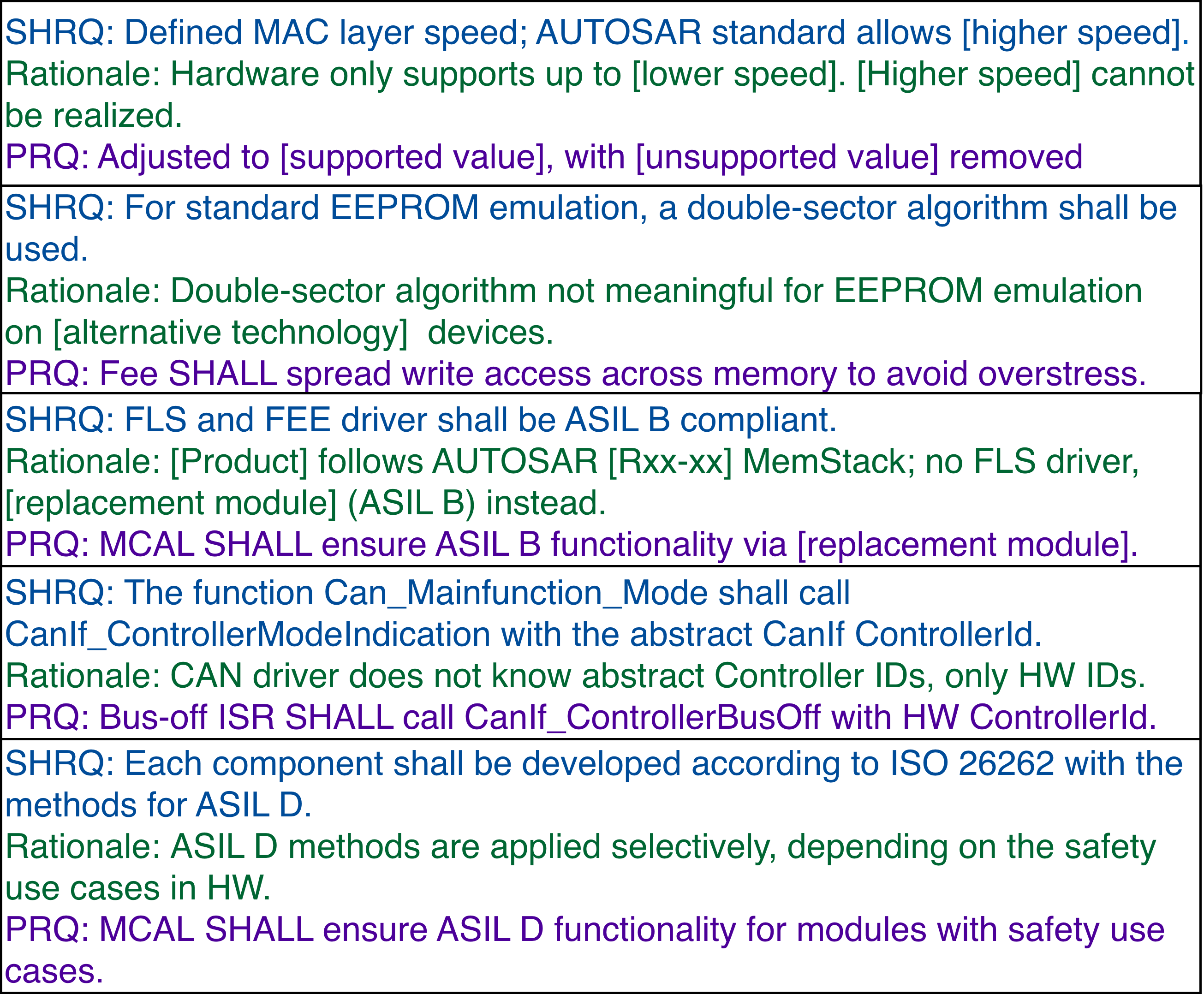}
\vspace{-3mm}
\caption{Examples of Approved with Deviation}
\label{fig: Example of Approved with Deviation(short)}
\vspace{-3mm}
\end{figure}

A closer inspection of the rationales reveals several recurring patterns that illustrate the unique character of automotive RE (Examples in Fig.~\ref{fig: Example of Approved with Deviation(short)}). Unlike enterprise or consumer software, automotive requirements are deeply shaped by hardware capabilities, the evolution of specifications, and the need to integrate with tightly coupled execution environments. 
For example, deviations frequently arise from hardware limitations. Some SHRQs prescribe parameter ranges or supported modes that exceed the capabilities of the hardware (the first SHRQ in Fig.~\ref{fig: Example of Approved with Deviation(short)}). In other cases, deviations arise from \textit{Tool and Standardization Issues}, such as overly broad or outdated standard. The second SHRQ from specification shows that unrealistic standard requirement, where in EEPROM emulation were replaced with alternative formulations more aligned with current hardware.
The third example (\textit{Redundancy and Partitioning}) shows the corresponding responsibility reassigned to a replacement module.
\textit{API behavior} also drives deviations. The fourth SHRQ prescribes the use of API (e.g., CanIf controller), which could not be implemented directly at the driver level, where only hardware IDs were available, leading to re-scoped PRQs. Finally, \textit{Functional Safety and ASIL Compliance} imposes selective application of standards. While SHRQs often demanded uniform ASIL D compliance, PRQs implemented ASIL levels selectively, depending on module criticality and hardware support. The prevalence of this outcome indicates that many SHRQs are neither outright acceptable nor wholly unsuitable, but require negotiation to reconcile intent with feasibility.

Taken together, these cases reveal that deviation is not simply a sign of requirement quality but a structural mechanism through which automotive software negotiates between specifications, hardware feasibility, safety certification, and supplier differentiation. 
Unlike in general software engineering, where requirements are shaped primarily by customers and development teams, automotive requirements emerge from negotiation across multiple actors—standards bodies, OEMs, hardware constraints, and suppliers—making acceptance with deviation an inherent feature of the domain rather than an exception.



Overall, SHRQs fail to achieve full approval for two main reasons. First, many are rejected due to mis-scoping, irrelevance, or intrinsic quality defects. Second, even approved requirements may conceal limitations when feasibility or architectural constraints force deviations. Both outcomes underscore the dual challenge of requirements validation in the automotive industry: early scoping and alignment with industry specifications are necessary to filter out irrelevant or defective inputs, while technical feasibility assessments are essential.

\begin{tcolorbox}
\textbf{Finding 3:} Rejections are primarily driven by scope mismatches, hardware limitations, and conflicts with standards or specifications. Requirements approved with deviation reflect systematic adaptations in which stakeholder intent is preserved but reorganized to align with architectural partitioning, hardware capabilities, and integration constraints in automotive software systems.

\end{tcolorbox}

\subsection{Mapping Complexity}
\subsubsection{RQ4: Factors Influencing the Complexity of SHRQ–PRQ Mappings} 

To investigate how SHRQs are refined into PRQs and what drives the resulting complexity, we analyze both the structural distribution of SHRQ–PRQ mappings and the factors associated with more complex refinement patterns.

\textit{Mapping Distributions and Structural Forms}
Overall, SHRQ–PRQ mappings are dominated by simple one-to-one relationships, but also include substantial cases of decomposition and consolidation (Table \ref{tab:mapping_stats}). From the stakeholder perspective (SHRQ $\rightarrow$ PRQ (decomposition)), each approved SHRQ maps to 1.93 PRQs on average, with a median of one. At least half of the SHRQs are realized through a single PRQ, while a quarter map to two or more PRQs. The distribution is long-tailed, with a maximum of 67 PRQs derived from a single SHRQ. From the product perspective (PRQ $\rightarrow$ SHRQ (consolidation)), each PRQ traces back to 1.32 SHRQs on average. Although one-to-one mappings are most common, rare many-to-one relationships occur, with some PRQs consolidating inputs from more than 20 SHRQs. These cases reflect the consolidation of overlapping or partially redundant stakeholder inputs to ensure consistency and avoid duplication at the product level. Together, these results show that while simple mappings dominate, many-to-many relationships are a fundamental aspect of industrial requirements refinement. One-to-many mappings (such as Fig. \ref{fig:Requirement Engineering Process}) reflect the decomposition of SHRQs that span multiple functional responsibilities, whereas many-to-one mappings reflect consolidation across heterogeneous stakeholder inputs.

\begin{table}[h!]
\centering
\caption{Mapping Statistics between SHRQs and PRQs}
\vspace{-0.3cm}
\label{tab:mapping_stats}
\resizebox{\linewidth}{!}{
\begin{tabular}{lcccccccc}
\toprule
Mapping Direction & Count & Mean & Std & Min & 25\% & 50\% & 75\% & Max \\
\midrule
SHRQ → PRQ & 3,688 & 1.93 & 2.91 & 1 & 1 & 1 & 2 & 67 \\
\quad Spec-derived & 3,270 & 1.74 & 2.30 & 1 & 1 & 1 & 2 & 67 \\
\quad Non-spec & 418 & 3.40 & 5.58 & 1 & 1 & 2 & 3 & 46 \\
\midrule
PRQ → SHRQ & 5,392 & 1.32 & 0.80 & 1 & 1 & 1 & 1 & 23 \\
\bottomrule
\end{tabular}}
\vspace{-3mm}
\end{table}

\begin{figure}[h!]
\centering
\includegraphics[width=0.8\columnwidth]{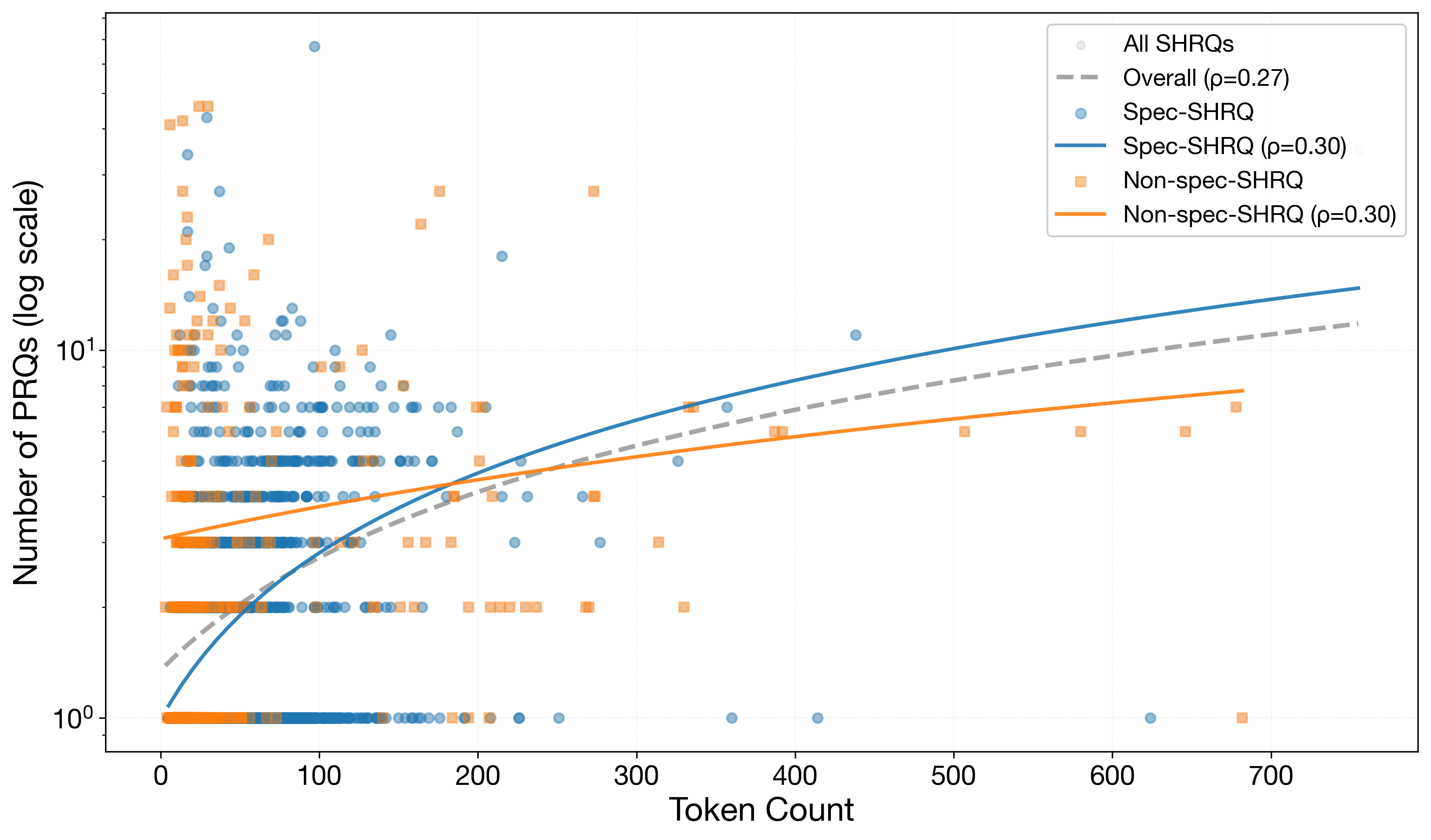}
\vspace{-0.5cm}
\caption{Mapping Complexity vs SHRQ Length: Overall and by Source}
\label{fig: Mapping Complexity vs SHRQ Length Overall and by Source}
\vspace{-0.3cm}
\end{figure}

\textit{Length}
To assess whether mapping complexity can be explained by requirement length, we analyze the relationship between SHRQ length and the number of mapped PRQs. Textual length shows a weak positive association with mapping
complexity. Using Spearman’s $\rho$, the token count ($\rho$ =0.27) exhibits
a statistically significant but weak correlation with the number of
mapped PRQs, indicating that while longer SHRQs tend to map to slightly more PRQs, the weak correlation suggests that textual length plays a minor role in mapping complexity. Notably, this pattern remains consistent across requirement sources: spec-SHRQs and non-spec SHRQs exhibit nearly identical weak correlations ($\rho$=0.30 and $\rho$=0.30, respectively). This consistency indicates that specification does not alter the relationship between textual length and mapping complexity. Fig.~\ref{fig: Mapping Complexity vs SHRQ Length Overall and by Source} shows substantial dispersion around the regression trend. SHRQs with similar token counts exhibit vastly different mapping patterns, ranging from simple one-to-one relationships to decompositions involving more than twenty PRQs. A short requirement such as “A safety management driver shall be provided” expands into 13 PRQs because it anchors a safety-critical mechanism spanning multiple modules, whereas a much longer SHRQ describing diagnostic faults across ASC/LIN/SPI channels results in few mappings through a single PRQ due to its localized architectural scope. These observations indicate that mapping complexity is driven less by textual properties than by the number of distinct functional and architectural concerns implicated by a requirement. Requirements spanning multiple subsystems, interfaces, or safety mechanisms require extensive decomposition, regardless of their textual brevity.

\textit{Role of Specification in Mapping Complexity}.
Requirement source further shapes mapping complexity. Specification-derived SHRQs map to fewer PRQs on average (mean $\approx 1.74$), reflecting their stable scope and alignment with established architectural and specification practices. In contrast, accepted non-spec SHRQs map to a larger number of PRQs on average (mean $\approx 3.40$), indicating that such requirements often require substantial decomposition, re-scoping, or clarification before implementation. Notably, the most extreme decomposition cases occur among specification-derived SHRQs (maximum of 67 PRQs). These cases typically arise when reusable specification requirements serve as reference points that must be instantiated and adapted across multiple product contexts. Thus, specification constrains refinement into predictable clusters in most cases, while also enabling highly complex decompositions when reuse and instantiation are required.

These findings show that SHRQ–PRQ mapping complexity is primarily driven by architectural scope, functional dispersion, and contextual assumptions rather than by surface textual characteristics. Structural mapping patterns—ranging from one-to-one to many-to-many relationships—provide concrete evidence of how these drivers manifest during industrial requirements refinement.

\begin{tcolorbox}
\textbf{Finding 4:} 
SHRQ–PRQ mapping complexity is only weakly related to textual length and is driven mainly by architectural scope and specification context. Most mappings are one-to-one, but many-to-many structures emerge when requirements span multiple components or safety mechanisms. While specifications generally stabilize refinement, they can increase complexity when reused across different product contexts.
\end{tcolorbox}

\subsection{Contextual Knowledge in Refinement}

\subsubsection{RQ5: Missing Contextual information in Refinement}

To address RQ5, we analyze what contextual information is missing in SHRQs and must be reconstructed during PRQ elaboration. Because a single requirement may lack multiple types of contextual information, we formulate this analysis as a multi-label classification problem. Domain engineers with experience in automotive requirements refinement defined a taxonomy of contextual information categories that reflect recurring information gaps observed in industrial practice. The final taxonomy comprises 13 categories, including \textit{operational context}, \textit{conditional logic}, \textit{functional behavior}, \textit{error handling}, \textit{configuration}, \textit{interface specification}, and \textit{safety and compliance}, among others. Using this domain-defined taxonomy, we employ a large language model (GPT-4o) to classify each SHRQ–PRQ pair by identifying which contextual information categories are absent from the SHRQ but explicitly introduced in the corresponding PRQ. Consistent with the approach used in RQ3, we conduct human-in-the-loop validation, in which domain experts review a stratified sample of the classifications and resolve disagreements by consensus. 


\textit{Distribution of Missing Contextual Information}
Our SHRQ-level analysis shows that missing contextual information is both widespread and systematic. When aggregating missing categories across all PRQs linked to each SHRQ, \textit{operational context} emerges as the most frequently absent category, missing in 81.5\% of SHRQs. This indicates that most SHRQs do not specify the system states, operating modes, or conditions under which they apply. Several other categories are also commonly missing. \textit{Conditional logic} is absent in 51.5\% of SHRQs, and \textit{functional behavior} in 45.0\%, suggesting that requirements often express desired outcomes without defining triggering conditions or detailed system responses. \textit{Error handling} (36.5\%) and \textit{configuration information} (29.5\%) are also frequently missing, highlighting the gap between stakeholder intent and implementation-level precision. Fig. \ref{fig:Requirement Engineering Process} shows a representative SHRQ–PRQ pair where multiple contextual dimensions, including implementation details, functional behavior, operational context, and conditional logic, are absent from the SHRQ and introduced during product-level elaboration. Overall, the five most common categories, each missing in more than a quarter of SHRQs, indicate that refinement typically involves enriching a recurring set of contextual dimensions rather than addressing isolated gaps.

\textit{Extent of Contextual Incompleteness per SHRQ}
At the individual requirement level, missing context is common. Among 3,688 unique SHRQs, each requirement is missing an average of 3.65 contextual categories, with a median of 3. Only 0.2\% are fully specified, while over 70\% are missing three or more types of context. The most common case is three missing categories, and more than a quarter lack five or more, showing that refinement usually requires reconstructing a large amount of implicit knowledge.

\textit{Co-occurrence of Missing Contextual Categories}
Missing contextual information usually appears in clusters rather than alone. Co-occurrence analysis shows that \textit{operational context} is the most central missing category, frequently appearing alongside others. The most common pairings combine \textit{operational context} with \textit{conditional logic}, \textit{functional behavior}, and \textit{error handling}. This suggests that refinement typically requires coordinated additions across several interdependent dimensions. Engineers often must reconstruct execution conditions, system behavior, and failure handling together rather than filling a single gap.

\begin{figure}[h!]
\centering
\includegraphics[width=0.75\columnwidth]{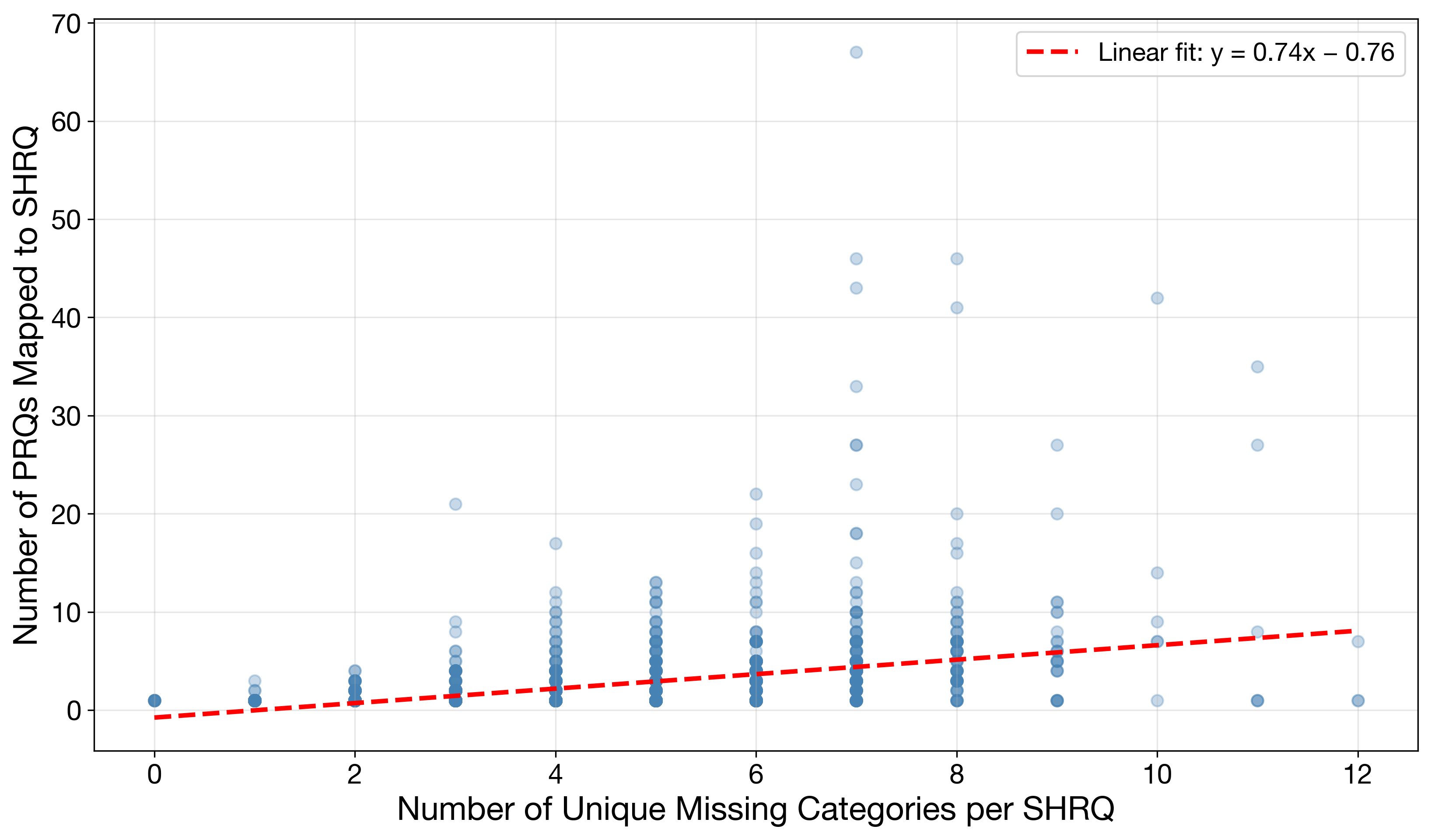}
\vspace{-0.3cm}
\caption{Missing Categories vs Number of Mapped PRQs}
\label{fig: RQ6_categories_vs_prq_boxplot}
\vspace{-0.5cm}
\end{figure}

\textit{Relationship Between Missing Context and Refinement Complexity}
To quantify this effect, we examine the relationship between the number of missing contextual categories in an SHRQ and the number of PRQs derived from it. Both Pearson (r=0.44, p<0.001) and Spearman ($\rho$=0.57, p<0.001) correlations show a moderate positive association, indicating that greater contextual incompleteness corresponds to higher refinement complexity. As shown in Fig.~\ref{fig: RQ6_categories_vs_prq_boxplot}, the relationship is monotonic with a stepwise pattern. SHRQs missing zero or one category almost always map to a single PRQ, while those missing four to five categories typically require two PRQs. When eight or more categories are missing, a single SHRQ expands into nearly seven PRQs on average, with some cases exceeding ten. SHRQs missing only one or two categories never result in highly fragmented mappings, indicating that extensive decomposition occurs only when substantial contextual information is absent.

\begin{tcolorbox}
\textbf{Finding 5:} 
SHRQs omit several interdependent types of contextual information, like operational context, conditional logic, and functional behavior, which must be reconstructed during product-level elaboration. These gaps rarely appear in isolation and strongly influence refinement effort, as SHRQs missing more contextual dimensions are much more likely to be decomposed into multiple PRQs.
\end{tcolorbox}

\section{Discussion}
\label{chapter/5_discussion}

\subsection{Synthesis: From Intake Filtering to Context Reconstruction}
Taken together, our findings suggest a two-stage picture of industrial
requirements refinement. Intake acts as a \textit{scope filter}: whether
an SHRQ enters refinement is decided largely by its origin and
responsibility boundaries, rarely by how well it is written (Findings~2
and~3). Refinement itself is \textit{knowledge reconstruction}: effort
is governed not by requirement length but by missing context and
architectural scope (Findings~4 and~5). Industry specifications link
the two stages: specification-derived SHRQs pass the filter more often
(83\% vs.\ 10\%) and decompose less (1.74 vs.\ 3.40 PRQs), plausibly
because standards pre-answer the context that engineers must otherwise
reconstruct---and where a specification is reused across products, the
same mechanism produces the most extreme decompositions (up to 67 PRQs).

\subsection{Implications for Practice}


First, early intake validation should explicitly consider the requirement source, scope, and standard alignment. Findings 2 and 3 show that non-specification SHRQs are often rejected or approved with deviation because they fall outside project scope, lack sufficient context, or conflict with architectural constraints. This extends prior work~\cite{femmer2014rapid, femmer2017rapid, fantechi2018requirement, farfeleder2011dodt} on requirement quality and smells, which focuses primarily on linguistic clarity and completeness, by showing that in automotive settings, acceptance depends more on scope and standard conformance than on surface-level textual quality. Lightweight intake checks that prioritize applicability, scope, and standards alignment, rather than linguistic polish alone, could substantially reduce downstream rework.

Second, deviation should be treated as a first-class artifact rather than an exception. Finding 3 shows that ``approved with deviation'' is not an edge case but a systematic result of aligning standardized requirements with hardware feasibility and integration constraints. However, current tools often capture deviations informally or not at all. Explicit deviation modeling that records rationale, constraints, and downstream impact would improve transparency, traceability, and auditability in safety-critical development.

Third, traceability management must move beyond link maintenance to actively support contextual enrichment. Findings 4 and 5 show that mapping complexity is only weakly associated with textual characteristics but strongly driven by missing contextual information and architectural scope. Tool support should therefore integrate standards references, hardware specifications, and module-level documentation directly into the refinement workflow. This would enable engineers to capture contextual knowledge when it is introduced, rather than reconstructing it later in an ad hoc way.

\subsection{Implications for Research and Future Work}

\textit{Beyond linguistic quality models.}
Findings 1 and 2 indicate that linguistic clarity alone is a poor predictor of acceptance. This contrasts with a large body of prior work on requirement smells and NLP-based quality assessment, which primarily targets ambiguity, verbosity, and syntactic defects (e.g., \cite{femmer2014rapid, femmer2017rapid, farfeleder2011dodt}). Our results suggest that such approaches, while valuable, are insufficient in isolation for safety-critical automotive systems, where feasibility, architectural scope, and standard compliance play a decisive role. Future research should combine NLP-based quality assessment with domain-aware checks that consider architectural scope, standard compliance, and feasibility constraints.

\textit{Modeling deviation and feasibility explicitly.}
The prevalence of approval with deviation (Finding 3) suggests the need for formal models that treat deviation as a structured refinement outcome rather than a binary failure. This opens opportunities to study deviation patterns, predict feasibility issues early, and support trade-off analysis in safety-critical RE.

\textit{Context-aware automation for requirement refinement.}
Findings 4 and 5 show that refinement is a knowledge-intensive process, relying on external artifacts like industry specifications and hardware documentation. Future work can explore techniques that recommend contextual sources, suggest missing requirement elements, or support decomposition based on architectural knowledge—moving beyond text similarity toward context-aware assistance.

\section{Related Work}
\label{chapter/6_related_work}

Prior work has highlighted the challenges posed by heterogeneous functions~\cite{broy2006challenges}, large-scale reuse~\cite{haghighatkhah2017improving, pretschner2007software}, and distributed development contexts~\cite{haghighatkhah2017improving, pretschner2007software}. Architectural frameworks such as AUTOSAR~\cite{AUTOSAR_Standard} promote modularity and reuse through standardized interfaces, which in turn shape how requirements are formulated in practice to remain compatible with established services, data types, and communication mechanisms~\cite{martinez2015survey, staron2021automotive, haghighatkhah2017automotive}.

Also, functional safety standards such as ISO 26262~\cite{ISO26262-2018} mandate rigorous requirements processes, including completeness, consistency, and bidirectional traceability across abstraction levels~\cite{ISO26262-2018, schneider2013literature}. Empirical studies~\cite{braun2014guiding, vogelsang2015model} report that industrial automotive requirements frequently exhibit inconsistencies, ambiguity, and limited verifiability despite these standards. While these works establish the importance of requirements engineering in automotive systems, they focus on processes and compliance, offering limited insight into how requirements evolve across abstraction levels in practice.

Against this domain background, a substantial body of research has examined the quality of natural-language requirements more broadly. Early frameworks such as IEEE 830~\cite{IEEE830-1998} and subsequent quality models emphasize properties including completeness, consistency, defects, and unambiguity~\cite{juergens2010can, pohl1996requirements, nuseibeh2000requirements, falessi2011empirical, kof2007treatment}. Studies on requirement smells, controlled vocabularies, and structured templates demonstrate that linguistic regularization can reduce ambiguity and improve readability~\cite{femmer2014rapid, femmer2017rapid, gilb2005competitive, fernandez2017naming,fantechi2018requirement, lami2004automatic}. More recent work~\cite{ferrari2017pure, femmer2025description, dalpiaznatural} applies natural language processing and computational linguistics techniques to analyze sentence structure, lexical density, and domain terminology at scale. These approaches provide valuable methods for identifying quality defects, but they typically treat requirements as isolated textual artifacts, without considering their role within a larger refinement and acceptance process or the constraints imposed by safety-critical domains.

Beyond individual requirement statements, traceability has long been recognized as essential for managing complexity in large software systems~\cite{goguen1993techniques, ramesh2002toward}. Extensive research has explored automated traceability recovery using information retrieval, machine learning, and deep learning techniques~\cite{cleland2012software, cleland2014software, borg2014recovering}. Empirical studies have also examined traceability practices in industry, highlighting challenges related to scalability and maintenance effort~\cite{winkler2010survey, cleland2014software}. Other work proposes modeling languages or refinement frameworks to support the transformation of informal stakeholder requirements into more formal specifications~\cite{li2015stakeholder, zave1997four}. 

Current work emphasizes creating or recovering traceability links rather than analyzing them as evidence of refinement decisions. As a result, little is known about why stakeholder requirements are accepted, rejected, consolidated, or approved with deviation, how mapping structures arise, or what drives refinement complexity in large-scale settings. This study treats stakeholder-to-product refinement as an empirical decision-making process and analyzes acceptance, rejection, and deviation.

\section{Conclusion}
\label{chapter/7_conclusion}
This paper examines how stakeholder-level requirements are refined into product-level requirements in industrial automotive software development and highlights the practical challenges of this transition. Using a large industrial dataset from Infineon, the study analyzes how requirements are evaluated, transformed, and contextualized in settings shaped by safety standards, architectural constraints, and hardware dependencies. The results reveal clear and systematic differences between stakeholder and product requirements in structure, granularity, and contextual completeness. The analysis further shows that decisions to accept, reject, or approve requirements with deviation are driven mainly by architectural scope, technical feasibility, and missing domain context, rather than by surface-level linguistic features. The observed stakeholder–product requirement mappings indicate that refinement complexity often arises from implicit assumptions that engineers must reconstruct from standards, specifications,  and hardware documentation. This explains why even concise stakeholder requirements can require substantial elaboration and why deviations are a routine part of industrial requirements intake. Overall, the study provides the first industrial-scale empirical characterization of the structural relationship between stakeholder and product requirements and the decision-making processes involved in refinement within the automotive industry. By linking RE theory with standards-driven industrial practice, the findings offer practical guidance for improving requirement refinement. Future work can build on these results by enabling earlier estimation of refinement and risk and by developing tool support for context-aware requirement elaboration.


\section{Data Availability Statement}
The industrial dataset used in this study is proprietary and subject to confidentiality restrictions that preclude public release. More information, including detailed results and taxonomy prompts, can be found on our website: \textbf{\underline{\url{https://sites.google.com/view/shrq2prq}}}.

\bibliographystyle{ACM-Reference-Format}
\bibliography{main}

\end{document}